\documentclass[sigconf]{acmart}

\copyrightyear{2025}
\acmYear{2025}
\setcopyright{acmlicensed}
\acmConference[CIKM '25] {Proceedings of the 34th ACM International Conference on Information and Knowledge Management}{ November 10--14, 2025}{Seoul, Republic of Korea.}
\acmBooktitle{Proceedings of the 34th ACM International Conference on Information and Knowledge Management (CIKM '25), November 10--14, 2025, Seoul, Republic of Korea}
\acmISBN{979-8-4007-2040-6/2025/11}
\acmDOI{10.1145/3746252.376087}

\usepackage{multirow}

\begin{document}

\title{Social Relation Meets Recommendation: Augmentation and Alignment}

\author{Lin Wang}
\affiliation{%
  \institution{The Hong Kong Polytechnic University}
  \city{Hong Kong}
  \country{China}}
\email{comp-lin.wang@connect.polyu.hk}

\author{Weisong Wang}
\affiliation{%
  \institution{Baidu Inc.}
  \city{Beijing}
  \country{China}}
\email{wangweisong@baidu.com}

\author{Xuanji Xiao\textsuperscript{*}}
\affiliation{%
  \institution{Shopee Inc.}
  \city{Shenzhen}
  \country{China}}
\email{growj@126.com}

\author{Qing Li\textsuperscript{*}}
\thanks{\textsuperscript{*}Corresponding author}
\affiliation{%
  \institution{The Hong Kong Polytechnic University}
  \city{Hong Kong}
  \country{China}}
\email{csqli@comp.polyu.edu.hk}






\renewcommand{\shortauthors}{Wang et al.}


\begin{abstract}

Recommender systems are essential for modern content platforms, yet traditional behavior-based models often struggle with cold users who have limited interaction data. Engaging these users is crucial for platform growth. To bridge this gap, we propose leveraging the social-relation graph to enrich interest representations from behavior-based models. However, extracting value from social graphs is challenging due to relation noise and cross-domain inconsistency.
To address the noise propagation and obtain accurate social interest, we employ a dual-view denoising strategy, employing low-rank SVD to the user-item interaction matrix for a denoised social graph and contrastive learning to align the original and reconstructed social graphs. Addressing the interest inconsistency between social and behavioral interests, we adopt a ”mutual distillation” technique to isolate the original interests into aligned social/behavior interests and social/behavior specific interests, maximizing the utility of both. 
Experimental results on widely adopted industry datasets verify the method's effectiveness, particularly for cold users, offering a fresh perspective for future research.
The implementation can be accessed at \url{https://github.com/WANGLin0126/CLSRec}.
\end{abstract}





\maketitle

\begin{figure}[tbp]
\vspace{-5pt}
\centering 
\includegraphics[height=0.15\textheight,width=\columnwidth]
{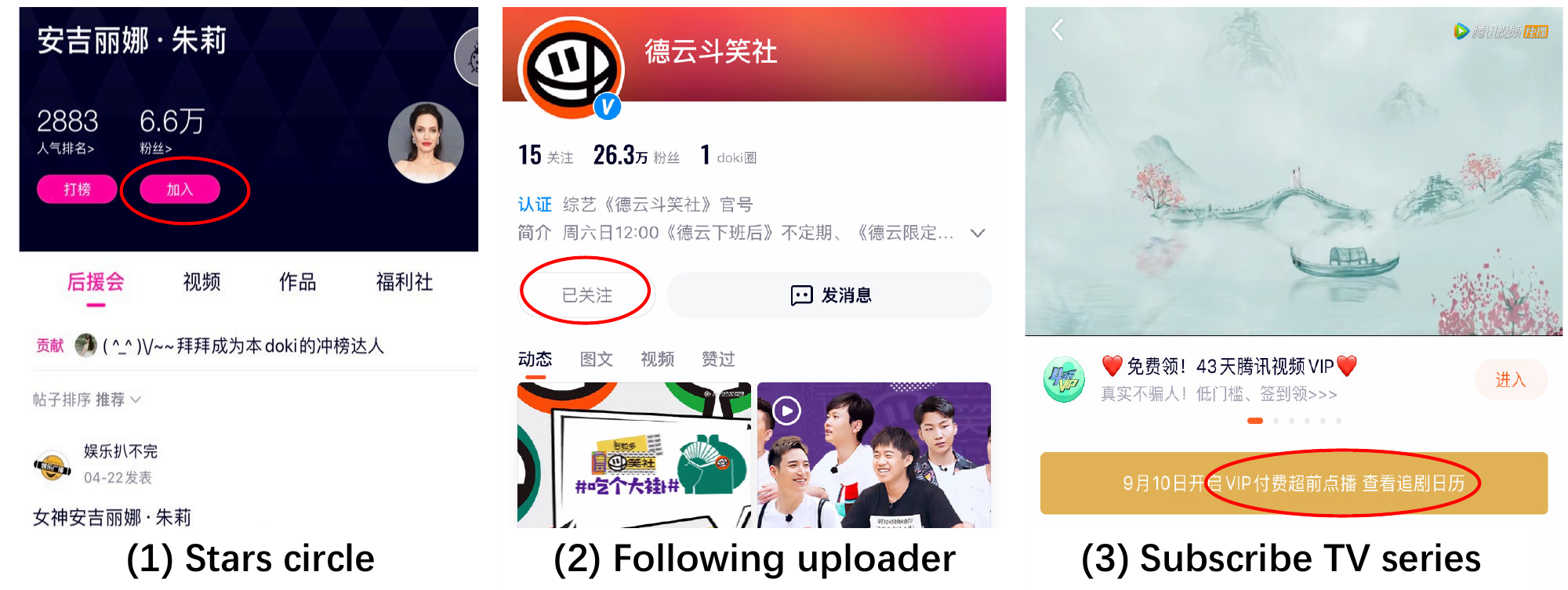}
\caption{Social interest information on a typical video platform encompasses: (1) tracking stars; (2) keeping up with uploaders; (3) subscribing to TV series.} 
\label{fig.videoplatform} 
\vspace{-15pt}
\end{figure}

\section{Introduction}
Recommender systems (RS) have now become core parts of major Internet platforms, providing personalized content to improve the user experience and drive engagement.
Traditional recommendation models\cite{cheng2016wide,zhou2018deep,li2023stan,ouyang2023click,lin2019pareto}, and typical content recommendation models\cite{covington2016deep,{xiao2024lottery4cvr},xiao2023neighbor,zhang2023modeling,ouyang2023knowledgesoftintegrationmultimodal}, score candidate items based on user characteristics extracted from behaviors (e.g., clicking, viewing). 
However, behavior-based recommendation models typically fail on \textbf{cold users}, who exhibit little behavior or are new to the platform~\cite{gope2017survey,sethi2021cold,bobadilla2012collaborative,lu2020meta,zhang2019contextual,li2017user}. 
Improving the personalized experience of \textit{cold users} on RS can be helpful to drive engagement, and converting 
\textbf{cold users} into \textit{active users} is vital to the continuous development of all content platforms.

We propose leveraging social graphs for joint social-behavior modeling, shifting beyond behavior-centric approaches. Social graphs capture diverse user relations—friendships, shared interests, and subscriptions—across platforms like YouTube, LastFM, and TikTok. Such relations often imply similar preferences, making social graph integration highly beneficial for recommendation systems. However, behavior-based models still struggle to effectively utilize this rich information.

\textbf{Social Relation Noise.} The heterogeneity of graph data introduces noise propagation, making the social interest inaccurate~\cite{xiao2023incorporating}. For example, users who are friends tend to have higher similarity, and their information propagation is considered beneficial for the performance of recommendation systems. However, 
the information propagation between celebrities and their fans needs to be carefully considered to avoid degrading the performance. On the other hand, considering the sparsity of social data, there is a significant proportion of users who have limited social connections. To enhance the representation capability of this sub-data in social relation, we adopt a \textbf{dual-view denoising} approach. The approach firstpplies low-rank SVD to the user-item matrix to obtain denoised user embeddings for a reconstructed social graph.
Then, contrastive learning between original and reconstructed social graphs is adopted as data augmentation for users, alleviating graph noise.

\textbf{Interest Inconsistency}. Although social relations provide numerous information for each user, there exists distinct interest inconsistency between the social interest representation and the behavioral interest representation~\cite{xiao2023incorporating}, which exposes challenge to effectively fuse the knowledge existing in the two interest. We adopt a "\textbf{mutual distillation}" method to co-extract the interest representations of social graphs and interactive behaviors. From the view of social interest, we distill the information from the social graph that is useful to the user-interest representation. From the view of behavior interest, we isolate two distinct interest representations from the behavior-based interest, i.e., social-interest similar representation and behavior-specific representation, to avoid knowledge loss in behaviors. In this way, we fully fuse the knowledge of two domains for recommendation.

In general, our contributions can be summarized as follows:
\begin{itemize}
\item We introduce an innovative social graph augmentation method that enhances conventional behavior-driven recommendation models through holistic user interest modeling.
\item The method employs dual-view contrast learning to denoise original social relations and "mutual distillation" to co-extract knowledge from both behaviors and social relations for maximal knowledge fusing, offering a promising fresh paradigm.
\item Experiments on widely adopted industry datasets demonstrate the effectiveness of the approach, particularly for cold users, highlighting its potential for broader adoption in social-behavior recommendation.
\end{itemize}

\section{Methodology}
In this section, we present our proposed \underline{C}ontrastive \underline{L}earning \underline{S}ocial \underline{Rec}ommendation (\textbf{CLSRec}). CLSRec first employs GNNs to learn user–item interaction embeddings, then filters noisy connections to reconstruct a denoised social graph for refined user representations. To enrich these representations, contrastive learning integrates embeddings from both the original and reconstructed graphs, capturing complementary information. Finally, CLSRec disentangles consistent and inconsistent signals between social interests and private behaviors, maximizing the utility of both for comprehensive user modeling.

\begin{figure}
    \centering
    \includegraphics[width=1.0\linewidth]{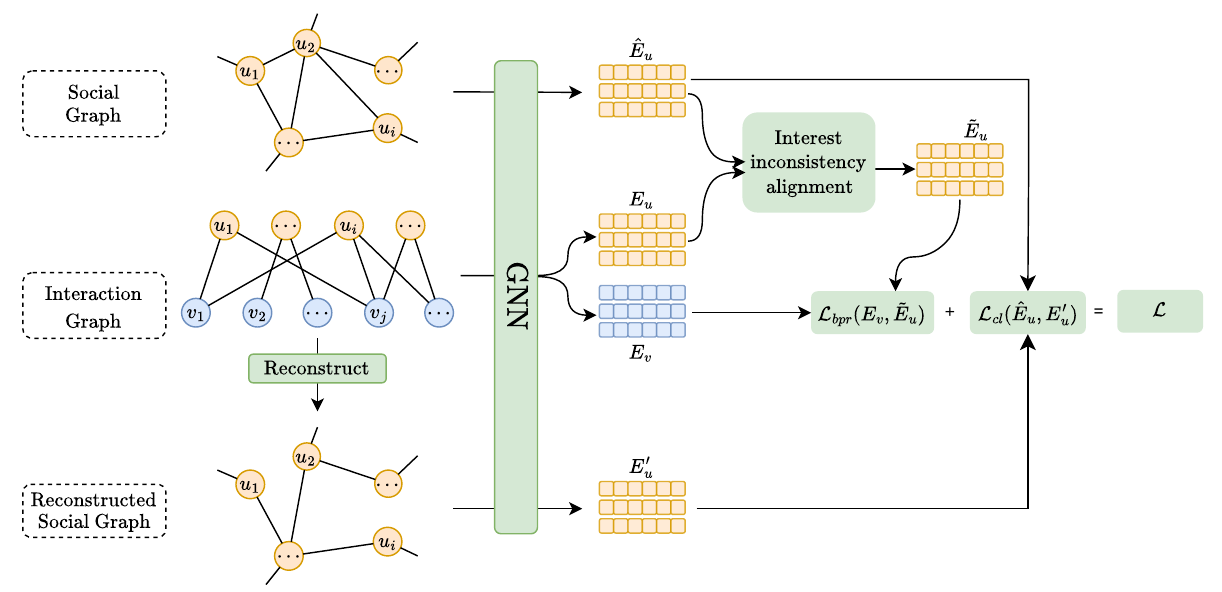}
    \caption{Our proposed contrastive learning social recommendation (\textbf{CLSRec}) framework.}
    \label{fig:CLSRec}
\end{figure}

\subsection{Information Extraction of Dual-view}
Social recommendation leverage two primary types of graphs to derive user and item embeddings: the user-item interaction graph and the user-user social graph. 
Specifically, the propagation for user-item interaction graph is modeled as follows:
\begin{align}
    e^{(l)}_{u,i}  = \sum_{j \in \mathcal{N}^{v}_{i}} \frac{1}{ \sqrt{|\mathcal{N}^{v}_i|} \sqrt{|\mathcal{N}^{u}_j|}} e^{(l-1)}_{v,j}, 
    e^{(l)}_{v,i}  = \sum_{j \in \mathcal{N}^{u}_{i}} \frac{1}{ \sqrt{|\mathcal{N}^{u}_i|} \sqrt{|\mathcal{N}^{v}_j|}} e^{(l-1)}_{u,j},
\end{align}
where $e^{l}_{u,i}$ and $e^{l}_{v,i}$ denotes the $i$-th user's and item's embedding in $l$-th layer, respectively. $\mathcal{N}^{v}_i$ is the item neighborhood of user $i$, while $\mathcal{N}^{u}_j$ are the user neighborhoods of the $j$-th item.
For the social graph, the layer-wise propagation could represented as,
\begin{align}
    \hat{e}^{(l)}_{u,i}  = \sum_{j \in \mathcal{N}^{u}_{i}} \frac{1}{ \sqrt{|\mathcal{N}^{u}_i|} \sqrt{|\mathcal{N}^{u}_j|}} e^{(l-1)}_{u,j},
\end{align}
where $\hat{e}^{(l)}_{u,i}$ denotes the $i$-th user embedding in $l$-th layer obtained from the social graph.
$\mathcal{N}^{u}_j$ is the user neighborhood of $j$-th user in the social graph.

After $L$  rounds of propagation, the final embeddings for users and items are computed by mean each layer embeddings, encapsulating the aggregated information from both the user-item interaction graph and the social graph.
\begin{align}
    E_u = \frac{1}{L+1}\sum_{l=0}^{L} {E}^{(l)}_{u},
    E_v = \frac{1}{L+1}\sum_{l=0}^{L} {E}^{(l)}_{v},
    \hat{E}_u = \frac{1}{L+1}\sum_{l=0}^{L} \hat{E}^{(l)}_{u}.
\end{align}

\subsection{Refine the Social knowledge}
Social graphs capture interest similarity (e.g., friends with shared preferences) but also contain noisy or weak ties that harm recommendation quality. To mitigate this, we reconstruct a refined social graph using user embeddings derived from the interaction graph via Singular Value Decomposition (SVD). Contrastive learning is then applied to align embeddings from the original and reconstructed graphs,  improving recommendation performance.
\subsubsection{Reconstructed Social View}
To construct the refined social graph view, SVD~\cite{abdi2007singular, cai2023lightgcl} is first utilized to decompose the user-item interaction matrix $A \in \mathbb{R}^{M \times N}$ as $A = USV^{\top}$. Here, $S \in \mathbb{R}^{M \times N}$ is a diagonal matrix of singular values, while $U \in \mathbb{R}^{M \times M}$ and $V \in \mathbb{R}^{N \times N}$ are orthogonal matrices of eigenvectors. The dimensions $M$ and $N$ represent the number of users and items, respectively. Singular values reflect the significance of their corresponding eigenvectors, referred to as principal components. To construct a low-rank approximation, we select the top $k$ singular values and their associated components, forming:
$A_k = U_kS_kV_k^{\top}$, where $A_k$ is a rank-$k$ approximation, $U_k \in \mathbb{R}^{M \times k}$ and $V_k \in \mathbb{R}^{N \times k}$ contain the top $k$ eigenvectors, and $S_k \in \mathbb{R}^{k \times k}$ is a diagonal matrix of the $k$ largest singular values. 
Then, a refined social view is reconstructed as,
\begin{align}
A^{S} = U_kS_kU_k^{\top},
\end{align}
where $A^{S}$ is a continuous approximation of the social view, differing from the binary nature of the original social and interaction graphs. For neighborhood aggregation on this reconstructed view, user embeddings are propagated as:
\begin{align}
E'^{(l)}_{u} = A^{S} \cdot E'^{(l-1)}_u,
\end{align}
where $l$ denotes the layer index. After $L$ propagation layers, the final user embeddings are obtained by taking the average across all layers, $E'_{u} = \frac{1}{L+1} \sum_{l=0}^{L} E'^{(l)}_{u}$.

\subsubsection{Contrastive Learning Alignment}
The reconstructed social view is considered an enhanced representation of social connections compared to the original social graph. For the same user from two different social graph view, the positive pairs is defined as \(\{(\hat{e}_{u,i}, e'_{u,i}) \mid i \in U\}\), while views of different users form negative pairs \(\{(\hat{e}_{u,i}, e'_{u,j}) \mid i, j \in U, i \neq j\}\)~\cite{wu2021self}. 
The reconstructed social view is derived by filtering user embeddings through SVD and computing the dot product. 
While this reconstructed view often differs significantly from the original social graph, we aim to ensure that similar users in the original graph remain similar, and dissimilar users remain dissimilar. To achieve this, we employ contrastive loss~\cite{chen2020simple, gutmann2010noise} to maximize agreement for positive pairs and minimize it for negative pairs:
\begin{align}
    \mathcal{L}_{cl} = \sum_{i \in U} - \mathrm{log}\frac{\mathrm{exp}(s(\hat{e}_{u,i}, e'_{u,i})/\tau)}{\sum_{j \in U} \mathrm{exp}(s(\hat{e}_{u,i}, e'_{u,j})/\tau)},
\end{align}
where $s(\cdot, \cdot)$ is the similarity function, set as cosine similarity in this work, and $\tau$ is a hyperparameter.

\subsection{Interest Inconsistency Alignment}
\begin{figure}[tbp] 
\vspace{-5pt}
\centering 
\includegraphics[width=0.9\columnwidth]{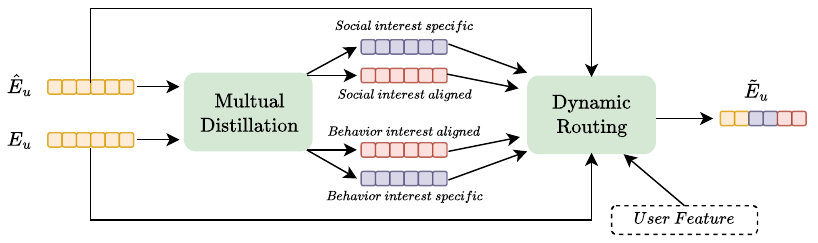}
\caption{Interest-inconsistency alignment by mutual distillation and dynamic routing.
} 
\label{fig.main2} 
\end{figure}

Although both social relations and user behaviors inform user interests, they may be inconsistent, and naive fusion risks introducing noise. To mitigate this, we propose a mutual distillation method that extracts complementary signals while isolating conflicts (Fig.~\ref{fig.main2}). Specifically, a co-attention mechanism compares behavior embeddings ($E_u$) and social embeddings ($\hat{E}_u$), generating weight vectors to decompose each into aligned and specific interests.
Inspired by recent works~\cite{Xiong2016DynamicCN,Si2023WhenSM} on knowledge alignment, we adopt the co-attention mechanism to align user interests. Specifically, we compute dual-weight vectors $W_s, W_i \in R^d$ as:
\begin{align}
    W_b &= \mathrm{softmax}(\mathrm{tanh}(\hat{E_u} W_b E_u)), \\
    W_s &= \mathrm{softmax}(\mathrm{tanh}(E_u W_s \hat{E}_u)),
\end{align}
where $W_b, W_s \in R^{d \times d}$ are learnable weight matrices, and $softmax$ normalizes the weights. The aligned weights $W_b$ and $W_s$ represent affinity scores for behavior and social interests, respectively. Elements in $E_u$ and $\hat{E}_u$ with higher weights in $W_b$ and $W_s$ are indicative of aligned interests, while those with lower weights capture specific interests. The isolating process is defined as:
\begin{align}
    E_{ua} &= \{ E_{uj} \mid W_{bj} \geq \gamma_i \}, \quad E_{uu} = \{ E_{uj} \mid W_{bj} < \gamma_i \}, \\
    \hat{E}_{ua} &= \{ \hat{E}_{uj} \mid W_{sj} \geq \gamma_s \}, \quad \hat{E}_{uu} = \{ \hat{E}_{uj} \mid W_{sj} < \gamma_s \},
\end{align}
where $E_{uj}, \hat{E}_{uj}, W_{bj}, W_{sj}$ denote the $j$-th elements of $E_u, \hat{E}_u, W_b, W_s$, and $\gamma_i, \gamma_s$ are thresholds.

This process preserves both shared and individual-specific interests, yielding three representations from behaviors ($E_u, E_{ua}, E_{uu}$) and socials ($\hat{E}u, \hat{E}{ua}, \hat{E}_{uu}$). To fuse them, we adopt a gated routing strategy, where the gate input ($user_fea$) encodes behavior frequency and social diversity, enabling users to dynamically emphasize the most informative representation. The routing is defined as:
\begin{align}
    G_i &= \mathrm{softmax}(\mathrm{MLP}(user\_fea)), \\
    \tilde{E}_u &= \sum_i G_i \cdot E_i,
\end{align}
where $E_i$ represents one of the six disentangled interest representations, $G_i$ is the learnable gating parameter, and $\tilde{E}_u$ is the final fused user representation combining behavioral and social information.

\begin{table*}[th]

\caption{Recommendation performance comparison. The best results are highlighted in bold. Improve denotes the relative improvement (\%) of our method over the best baseline.}
\label{tab:per}
\setlength{\tabcolsep}{2pt} 
\resizebox{1.0\textwidth}{!}{
\begin{tabular}{cc|ccccccc|ccccccc}
\hline
\multirow{2}{*}{Dataset} & \multirow{2}{*}{Metric} & \multicolumn{7}{c|}{Recommendation   performance}                                                                                        & \multicolumn{7}{c}{Cold Users}                                                                                            \\ \cline{3-16} 
                         &                         & BPR    & DiffNet & LightGCN & SocialLGN       & GBSR            & \begin{tabular}[c]{@{}c@{}}CLSRec\\      (Ours)\end{tabular} & Improve & BPR    & DiffNet & LightGCN & SocialLGN & GBSR   & \begin{tabular}[c]{@{}c@{}}CLSRec\\      (Ours)\end{tabular} & Improve \\ \hline
\multirow{6}{*}{LastFM}  & Precision@10            & 0.0922 & 0.1727  & 0.1961   & 0.1972          & 0.1979          & \textbf{0.2002}                                              & 1.16\%  & 0.0282 & 0.0417  & 0.0417   & 0.0458    & 0.0511 & \textbf{0.0583}                                              & 14.09\% \\
                         & Precision@20            & 0.0720 & 0.1215  & 0.1358   & 0.1368          & 0.1370          & \textbf{0.1384}                                              & 1.02\%  & 0.0209 & 0.0271  & 0.0313   & 0.0333    & 0.0361 & \textbf{0.0417}                                              & 15.51\% \\
                         & Recall@10               & 0.0962 & 0.1779  & 0.2003   & 0.2026          & 0.2030          & \textbf{0.2057}                                              & 1.33\%  & 0.1151 & 0.1713  & 0.1727   & 0.1974    & 0.2089 & \textbf{0.2353}                                              & 12.64\% \\
                         & Recall@20               & 0.1499 & 0.2488  & 0.2769   & 0.2794          & 0.2800          & \textbf{0.2829}                                              & 1.04\%  & 0.1615 & 0.2407  & 0.2416   & 0.2663    & 0.2690 & \textbf{0.3056}                                              & 13.61\% \\
                         & NDCG@10                 & 0.1099 & 0.2219  & 0.2536   & 0.2566          & 0.2573          & \textbf{0.2617}                                              & 1.71\%  & 0.0828 & 0.1107  & 0.1374   & 0.1419    & 0.1510 & \textbf{0.1624}                                              & 7.55\%  \\
                         & NDCG@20                 & 0.1321 & 0.2474  & 0.2788   & \textbf{0.2883} & \textbf{0.2885} & 0.2872                                                       & -0.45\% & 0.0989 & 0.1309  & 0.1560   & 0.1643    & 0.1730 & \textbf{0.1852}                                              & 7.05\%  \\ \hline
\multirow{6}{*}{Ciao}    & Precision@10            & 0.0145 & 0.0238  & 0.0271   & 0.0276          & 0.0280          & \textbf{0.0285}                                              & 1.79\%  & 0.0061 & 0.0104  & 0.0131   & 0.0134    & 0.0135 & \textbf{0.0140}                                              & 3.70\%  \\
                         & Precision@20            & 0.0111 & 0.0182  & 0.0202   & 0.0205          & 0.0201          & \textbf{0.0212}                                              & 3.41\%  & 0.0047 & 0.0081  & 0.0096   & 0.0097    & 0.0096 & \textbf{0.0101}                                              & 4.12\%  \\
                         & Recall@10               & 0.0220 & 0.0341  & 0.0410   & 0.0430          & 0.0431          & \textbf{0.0441}                                              & 2.32\%  & 0.0208 & 0.0339  & 0.0429   & 0.0441    & 0.0445 & \textbf{0.0458}                                              & 2.92\%  \\
                         & Recall@20               & 0.0339 & 0.0527  & 0.0591   & 0.0618          & 0.0622          & \textbf{0.0635}                                              & 2.09\%  & 0.0328 & 0.0539  & 0.0616   & 0.0630    & 0.0638 & \textbf{0.0648}                                              & 1.57\%  \\
                         & NDCG@10                 & 0.0229 & 0.0359  & 0.0437   & 0.0441          & 0.0448          & \textbf{0.0456}                                              & 1.79\%  & 0.0138 & 0.0248  & 0.0319   & 0.0328    & 0.0330 & \textbf{0.0337}                                              & 2.12\%  \\
                         & NDCG@20                 & 0.0260 & 0.0403  & 0.0478   & 0.0486          & 0.0488          & \textbf{0.0502}                                              & 2.87\%  & 0.0179 & 0.0316  & 0.0384   & 0.0394    & 0.0401 & \textbf{0.0405}                                              & 1.00\%  \\ \hline
\end{tabular}
}
\end{table*}

\subsection{Training and Optimization}
Ultimately, the dot product of these two embeddings $\tilde{E}_u$ and $E_v$ quantifies the user's predicted rating of the item. A higher score suggests an increased probability of the user selecting the product  $\hat{Y} = \tilde{E}_u^{\top}E_v$, where $\hat{Y} \in \mathbb{R}^{ M \times N}$, the element $\hat{y}_{ij}$ in which indicates the score of user $i$ rating item $j$.

For the recommendation, we use the Bayesian Personalized Ranking (BPR) \cite{rendle2012bpr} loss, which states,
\begin{align}
    \mathcal{L}_{bpr} = -\sum_{u=1}^{M}\sum_{i\in\mathcal{N}_{u}}\sum_{j \notin \mathcal{N}_{u}} \mathrm{ln} \sigma(\hat{y}_{ui} - \hat{y}_{uj}) + \lambda||E^{(0)}||^{2},
\end{align}
where $\lambda$ is a hyperparameter which controls the regularization strength and $E^{0} = [E^{(0)}_u, E^{(0)}_v]$ are the init value of user and item embedding.
Therefor, the final loss function is  $\mathcal{L} = \mathcal{L}_{bpr} + \alpha \mathcal{L}_{cl}$, where $\alpha$ is the hyperparameter to balance the BPR loss and contrastive learning loss.

\section{Experiments}

Our experiments are conducted on two real-world industry datasets, which vary in terms of scale and sparsity: LastFM\footnote{https://grouplens.org/datasets/hetrec-2011/} and Ciao\footnote{https://www.cse.msu.edu/~tangjili/datasetcode/truststudy.htm}. LastFM is a music-related dataset that records the friendships of 1,892 users and their interactions with 17,632 music items. Ciao is a database covering online shopping, containing ratings from 7,375 users on 105,114 products as well as the social relationships between users. 
During the experiments, we randomly split the interaction data of each dataset into a training set and a test set, with 80\% of the data used for training and the remaining 20\% used for testing. To fairly tune the hyperparameters and prevent overfitting, we randomly select 10\% of the training data as a validation set.


To evaluate the performance of \textbf{CLSRec}, we compare it with the following 4 state-of-the-art methods: BPR\cite{rendle2012bpr}, a classic pairwise collaborative filtering algorithm optimized by maximizing the difference between the probability of a user interacting with positive samples and negative samples.
DiffNet\cite{wu2019neural}, a graph-based social recommendation method that generates final user representations by integrating user representation vectors from both the user-item interaction graph and the social graph.
LightGCN\cite{he2020lightgcn}, a widely applied model based on NGCF that further simplifies the model structure.
SocialLGN\cite{liao2022sociallgn}, that is currently the most advanced social recommendation graph model based on LightGCN, and GBSR~\cite{yang2024graph}, which uses the information bottleneck principle to denoise social graphs and improve recommendation performance.

\subsection{Performance Comparison}

Table \ref{tab:per} compares the proposed method with state-of-the-art baselines on the LastFM and Ciao datasets using Precision, Recall, and NDCG metrics. On the LastFM dataset, the proposed method consistently outperforms GBSR in most metrics, achieving notable improvements in Precision@10, Recall@10, and NDCG@10, with gains of up to 1.71\%. On the Ciao dataset, the proposed method demonstrates significant improvements in Precision and NDCG, particularly at top-ranked positions, with gains of up to 3.41\%.

Additionally, for the cold-user, where users with fewer than 20 interactions~\cite{liao2022sociallgn}, CLSRec exhibits significant and substantial relative-gains across all metrics, with improvements exceeding \textbf{15\% in Precision, 13\% in Recall, and 7\% in NDCG} on LastFM dataset. 
On Ciao, the performance gains are more modest, with slight improvements in Recall and NDCG, while Precision shows mixed results depending on the recommendation list length. 
Overall, CLSRec effectively improves top-ranked recommendations and mitigates the cold-user problem, achieving substantial gains on LastFM and Ciao, validating its robustness across datasets.


\subsection{Ablation Study}
Table~\ref{tab:ablation_clsrec} shows the ablation study of CLSRec on the LastFM dataset. Removing the contrastive loss (CL) slightly degrades performance, indicating its role in enhancing representation learning. Further removing the interest inconsistency alignment (IIA) leads to a more noticeable drop, demonstrating its importance in capturing complementary information. Overall, both components contribute to the effectiveness of the proposed CLSRec.


\begin{table}[htbp]
\centering
\caption{Ablation Study of CLSRec on LastFM dataset}
\begin{tabular}{lccc}
\toprule
\textbf{Method} & \textbf{Precision@20} & \textbf{Recall@20} & \textbf{NDCG@20} \\
\midrule
CLSRec & 0.1384 & 0.2829 & 0.2872  \\
w/o CL & 0.1368  & 0.2794 & 0.2883  \\
w/o CL \& IIA & 0.1358 & 0.2769  & 0.2788  \\
\bottomrule
\end{tabular}
\label{tab:ablation_clsrec}
\end{table}
\section{Conclusion}
In this paper, we proposed a CLSRec recommendation method to fuse social relation into behavior-based recommendations, which leverages the reconstructed social view to enhance the user interest representation, addressing key challenges like social relation noise and interest inconsistency. 
Experiments validate the effectiveness of the approach and dramatically improve recommendations for cold users. The dual-view modeling and social-interest alignment method establishes a promising paradigm, suggesting potential for broad adoption in social-behavior recommendation for future research.

\begin{acks}
This project has been supported by the Hong Kong Research Grants Council under General Research Fund (project no. 15200023).
\end{acks}

\newpage\clearpage
\balance
\bibliographystyle{ieeetr}
\bibliography{references}

\end{document}